# NMR study of superconductivity and spin fluctuations in hole-doped superconductor Ca$_{1-x}$Na$_x$Fe$_2$As$_2$ ($T_c$ =32 K)*

**Ma Long(马龙)[1], Zhang Jin-Shan(张金珊)[2], Wang Du-Ming(王笃明)[1],**

**He Jun-Bao(何俊宝)[1]，Xia Tian-Long(夏天龙)[1]，Chen Gen-Fu(陈根富)[1],**

**Yu Wei-Qiang(于伟强)[1]\*\***

[1] *Department of Physics, Renmin University of China, Beijing 100872, China*

[2] *School of Energy, Power and Mechanical Engineering, North China Electric Power University,*

*Beijing 102206, China*

\* Yu Wei-qiang and Chen Gen-Fu are supported by the National Basic Research Program of China (the 973 program) and the NSFC. Yu Wei-qiang also acknowledges the support by Program for New Century Excellent Talents in University. Zhang Jin-Shan is supported by the NSFC and Fundamental Research Funds for the Central Universities. Xia Tian-Long is supported by the NSFC.

\*\* Email: wqyu_phy@ruc.edu.cn

(Received )

We report both $^{23}$Na and $^{75}$As NMR studies on hole-doped Ca$_{1-x}$Na$_x$Fe$_2$As$_2$ superconducting single crystals (x ≈0.67) with $T_c$ =32 K. Singlet superconductivity is suggested by a sharp drop of the Knight shift $^{75}K$ below $T_c$. The spin-lattice relaxation rate $1/T_1$ does not show the Slichter-Hebel coherence peak, which suggests an unconventional pairing. The penetration depth is estimated to be 0.24 $\mu$m at $T$=2 K. $1/^{75}T_1T$ shows an anisotropic behavior and a prominent low-temperature upturn, which indicates strong low-energy antiferromagnetic spin fluctuations and supports a magnetic origin of superconductivity.

The discovery of high-temperature superconductivity in iron pnictides[1-4] has attracted intensive research interests in the last several years comparing with cuprates. The undoped iron pnictides are usually antiferromagnetic (AFM) metals with multiple electron bands on the Fermi surface[5], while superconductivity is induced by the suppression of long-range antiferromagnetism upon electron or hole-doping[1-4]. The pairing symmetry and the spin fluctuations in iron pnictides with different structures are the research focus to understand the paring mechanism. Among iron pnictides, the $ThCr_2Si_2$ (122) structure family has been mostly studied due to the availability of single crystals. All parent compounds have a high antiferromagnetic transition temperature. Strong low-energy spin fluctuations have been found upon electron doping and suggest a magnetic origin of superconductivity[6,7].

Under pressure or upon hole doping, the 122 structure $CaFe_2As_2$ shows many unique properties in contrast to $BaFe_2As_2$ and $SrFe_2As_2$ . For example, bulk superconductivity has been found in $BaFe_2As_2$ and $SrFe_2As_2$ under pressure[8], whereas filamentary superconductivity is found in $CaFe_2As_2$ with very low pressure ($P \geq 3$ kbar)[9,10]. Later studies on $CaFe_2As_2$ revealed that superconductivity is only possible under non-hydrostatic pressure[11], right at the phase boundary between the low-pressure orthorhombic phase and the high-pressure collapsed tetragonal phase[12]. Upon hole doping, superconductivity has also been observed with $T_c \approx 30$ K. For $Ba_{1-x}K_xFe_2As_2$, the optimal doping is about $x \approx 0.4$[13]. For $Ca_{1-x}Na_xFe_2As_2$, the optimal doping (with $T_c \approx 33$ K) is reported as $x \approx 0.67$, which is very high[14,15]. Recently, theoretical studies propose that the pairing symmetry and the spin fluctuations behaviors in heavily hole-doped compounds are probably different from the underdoped ones[16,17], although conclusive experimental evidence is still lacking.

In this report, we present our NMR studies of the superconductivity and the spin fluctuations on single crystalline $Ca_{0.33}Na_{0.67}Fe_2As_2$ with $T_c \approx 32$ K, and compare it with other iron pnicides. Singlet superconductivity is suggested by a sharp drop of $^{75}$As and $^{23}$Na Knight shift ($^{23}K$ and $^{75}K$) below $T_c$. The absence of the coherence peak suggests that the pairing symmetry is not a conventional s-wave type. In the normal state, the spin-lattice relaxation rate $1/^{75}T_1$ is anisotropic with different field orientations. An obvious Curie-Weiss upturn develops in $1/^{75}T_1T$ as the temperature cools down toward $T_c$. The anisotropy and the upturn behavior in $1/^{75}T_1T$ are strong evidence for low-energy AFM spin fluctuations.

The $Ca_{0.33}Na_{0.67}Fe_2As_2$ single crystals were synthesized by flux-growth method[18]. Our NMR measurements were applied on crystals with dimensions of $3 \times 2 \times 0.1 mm^3$. In Fig. 1, We show the temperature dependence of the in-plane resistivity and the dc susceptibility of the $Ca_{0.33}Na_{0.67}Fe_2As_2$ single crystal. The abrupt sharp drop of the in-plane resistivity and the dc susceptibility at 32 K indicate the high homogeneity of the sample. The superconducting transition is also monitored *in situ* by the ac susceptibility measurements. As shown in Fig. 2(a), the superconducting transition is very sharp as indicated by a sharp decrease of the ac susceptibility at 32 K. We performed both $^{23}$Na (S=3/2) and $^{75}$As (S=3/2) NMR studies, with the magnetic field applied along the crystal (*ab*) plane or the *c-axis*. The $^{75}$As spectrum is broad with a full-width of half maximum (FWHM) of 300 kHz at 35 K (see Fig. 2(b)), whereas the $^{23}$Na has a narrow spectrum with a FWHM of 12 kHz at 40 K (see Fig. 5). The Knight shift is calculated by $K(T) = (f - \gamma_n B)/\gamma_n B$, where *f* is the central frequency of the spectra, $\gamma_n$ is the gyromagnetic ratio of the nucleus, and *B* is the external magnetic field. The spin-lattice relaxation rate $1/^{75}T_1$ is measured by the inversion-recovery method, where the spin-recovery of both nuclei follows the standard form $I(t)/I(0)=(1-a(0.1e^{-t/T_1} + 0.9e^{-6t/T_1}))$ for *S*=3/2 nuclei.

We first show the Knight shifts $^{75}K$ and $^{23}K$ in Fig. 3, with field applied along the *c*-axis. In the normal state, $^{75}K$ does not change much with temperature from 250 K down to $T_c$. The $^{23}K$ is much smaller than $^{75}K$, which is reasonable because $^{23}$Na is located between the FeAs planes and therefore has a weaker hyperfine coupling to the Fe moments. Just below $T_c$, both $^{75}K$ and $^{23}K$ drop sharply. Since the Knight shift $K(T)$ measures local static susceptibility, the decrease below $T_c$ indicates a loss of spin susceptibility. This is a strong evidence for singlet pairing. The singlet superconductivity is consistent with other iron pnictides, which supports a universal pairing mechanism and puts a strong constraint to the theory of superconductivity.

The singlet pairing limits the orbital pairing function to *s*-wave, *d*-wave or other even symmetry types. We further study the orbital symmetry from the spin-lattice relaxation rate $1/T_1$. The spin-lattice relaxation rate $1/T_1$ sums the weighted dynamic spin susceptibility, with $1/T_1T \propto \sum_q A_{hf}^2(q) \text{Im} \chi_\perp(q,\omega)/\omega$, where $A_{hf}(q)$ is the hyperfine coupling, $\chi_\perp(q)$ is the dynamical susceptibility perpendicular to the external field, and $\omega$ is the NMR frequency. For a conventional (or BCS-like) s-wave superconductor, $1/T_1$ usually shows a Hebel-Slichter coherence peak[20] just below $T_c$, and drops with an activation function far below $T_c$. For a d-wave superconductivity, the $1/T_1$ usually decreases below $T_c$ without coherence peak, and approaches a power-law $1/T_1 \sim T^3$ far below $T_c$.

The $1/^{75}T_1$ data of the $Ca_{0.23}Na_{0.67}Fe_2As_2$ crystal at different temperatures is shown in the main panel of Fig. 4. In the normal state, the $1/^{75}T_1$ decreases as temperature drops. Below $T_c$, the $1/^{75}T_1$ is highlighted by a log-log plot in the inset of Fig. 4. $1/^{75}T_1$ drops sharply just below $T_c$, whereas the Hebel-Slichter coherence peak is not seen. The absence of the coherence peak is not consistent with a conventional s-wave pairing symmetry. In literature, the coherence peak has not been seen in any Fe superconductor, regardless of sample qualities[21-28]. Such an unconventional behavior may be accounted by several scenarios in Fe superconductors. In many iron pnictides, ARPES studies shows multiple isotropic gaps with s-wave pairing symmetry[29,30], and an $s_\pm$ pairing has been proposed due to interband transitions[31,32]. The coherence peak for the $s_\pm$ superconductivity may be suppressed by interband impurity scatterings[33-37]. In heavily hole-doped $KFe_2As_2$ and the isovalent-doped $BaFe_2As_{2-x}P_x$, however, superconducting gaps with line nodes were proposed[7,38], to account for the absence of the coherence peak[7]. For the current compounds, our NMR data are not sufficient to determine a propriate scenario.

Below $T_c/2$ (16 K), the $1/^{75}T_1$ decreases with reducing temperature by $1/^{75}T_1 \propto T$. Such a low-power law behavior cannot be described by either an s-wave or a d-wave pairing. We think that the low power-law behavior may be affected by the vortex core contribution, and therefore does not reflect the pairing symmetry of the superconducting region.

Next we study the penetration depth determined from the NMR spectra below $T_c$. Theoretically[39], the in-plane penetration depth, $\lambda_{ab}$ can be estimated from local distribution of magnetic field with $H<<H_{c2}$, that is, $\lambda_{ab}^{-2} \approx \Delta B_{local}/0.0609\phi_0$, where $\Delta B_{local} = \Delta f/\gamma_n$, $\Delta f$ is the NMR linewidth, and $\phi_0$ is the quantum flux.

The narrow linewidth of $^{23}$Na spectra allows us to estimate the $\lambda_{ab}$. Fig. 5 shows the $^{23}$Na linewidth ($\propto \lambda_{ab}^{-2}$) at different temperatures obtained from Gaussian fit. The sharp increase below 30 K indicates the onset of superconductivity. At the lowest temperature, the NMR linewidth is about 35 kHz, from which the $\lambda_{ab}$ is estimated to be 0.24 $\mu$m. We note that the penetration depth is close to the value obtained in $Ba_{1-x}K_xFe_2As_2$ with a similar $T_c$[40].

We also look closely at the temperature behavior of the linewidth. With temperature decreasing below $T_c$, the linewidth first increases sharply with temperature, then tends to saturates at low temperatures. A conventional s-wave fit to the linewidth, as shown by the solid line in Fig. 5, does not work well, which may also suggest an unconventional orbital pairing symmetry.

Finally we focus on the normal state properties. For a Fermi liquid, the spin-lattice relaxation rate usually follows the Korringa behavior with a constant $1/T_1TK^2$. To compare with a Fermi liquid, we replot $1/^{75}T_1T$ at different temperatures in Fig. 6. $1/^{75}T_1T$ shows a prominent decrease with temperature down to $T_c$, whereas $K$ does not change much with temperature (see Fig. 3), which clearly deviates from canonical Fermi liquid behavior.

We fit the $1/T_1T$ data with a Curie-Weiss function $1/T_1T = A/(T + \Theta)$, as shown by the solid line in Fig. 6. The fitting gives a large value of $\Theta \approx 100 \pm 5K$, which suggests that the system is distant from magnetic ordering ($\Theta \leq 0$). Such low-temperature Curie-Weiss upturn in $1/^{75}T_1T$ is an evidence for 2D spin fluctuations in metals, proposed by T. Moriya et al.[41]. Furthermore, the fluctuations are consistent with an AFM type. First, the upturn behavior is not seen in $K(T)$ which measures spin fluctuations at $q = 0$. Since the $1/T_1T$ sums contributions from all momentum space, such behavior suggest strong antiferromagnetic spin fluctuations away from $q = 0$. Second, the $1/T_1$ is anisotropic with field orientations. As shown in Fig. 6, the spin-lattice relaxation rate is larger with field along the *ab*-plane than that along *c-axis*, with the anisotropic factor determined as $T_1^c/T_1^{ab} \approx 1.3$. Previous NMR studies on Fe superconductors[42,43] show that strong stripe AFM spin fluctuations lead to an anisotropic $1/T_1$ with $T_1^c/T_1^{ab} \approx 1.5$. Our slightly lower anisotropic factor indicates a similar, but weaker, AFM correlations. The existence of spin fluctuations supports a magnetic origin of superconductivity.

Our experimental evidences for singlet pairing and spin fluctuations are consistent with many electron and the hole-doped iron pnictides with lower dopings[6,7,42,44-47]. It is surprising, however, to see prominent low-energy spin fluctuations under such a large nominal doping (0.33 hole/Fe). Careful EDX measurements on crystals from the same batch confirm the high doping level[18]. In heavily hole-doped KFe$_2$As$_2$ (0.5 hole/Fe), the anisotropy and the low-temperature upturn in $1/^{75}T_1T$ were also reported, indicating strong AFM spin fluctuations[25]. For the electron-doped case, such as Ba(Fe$_{1-x}$Co$_x$)$_2$As$_2$, however, the low-energy spin fluctuations are very weak at dopings above $0.12e$/Fe[6]. This indicates an electron-hole asymmetry in the spin fluctuations of the doped iron pnictides. In fact, the superconducting phase diagram is also asymmetric upon electron doping and hole doping. The superconductivity persists up to much higher hole doping (0.5 hole/Fe in KFe$_2$As$_2$[13] and 0.5 hole/Fe in NaFe$_2$As$_2$[48]) than that of the electron doping (0.15 $e$/Fe for Ba(Fe$_{1-x}$Co$_x$)$_2$As$_2$[49] and 0.2 $e$/Fe for Ca(Fe$_{1-x}$Co$_x$)$_2$As$_2$[50]). Our NMR evidence of contrasting low-energy spin fluctuations may further explain the asymmetry of the superconducting phase diagram in the 122 class with a magnetic origin of superconductivity.

To summarize, our NMR Knight shift data in Ca$_{0.33}$Na$_{0.67}$Fe$_2$As$_2$ suggests singlet superconductivity. However, the temperature behaviors of the spin-lattice relaxation rate and the penetration depth are not consistent with a conventional $s$-wave superconductor. Prominent antiferromagnetic low-energy spin fluctuations are also seen by the the spin-lattice relaxation rate in the normal state, which supports a magnetic origin of superconductivity. The existence of low-energy spin fluctuations are consistent with similar heavily hole-doped Ba$_{1-x}$K$_x$Fe$_2$As$_2$, but in contrast to the heavily electron-doped Ba(Fe$_{1-x}$Co$_x$)$_2$As$_2$.

**Figures Caption**

Fig.1: (color online) Main panel: Temperature dependence of the in-plane resistivity of the $Ca_{0.33}Na_{0.67}Fe_2As_2$ single crystal. Inset: dc susceptibility vs temperature with a 10 Oe magnetic field applied along $c$-axis of the single crystal.

Fig. 2: (color online) (a) The resonance frequency $f$ of the NMR coil with the $Ca_{0.33}Na_{0.67}Fe_2As_2$ crystal inside. The change of $f$ is related to the ac susceptibility of the sample by $\Delta f / f \propto \chi_{ac}(T)$. The superconducting transition is indicated by a sharp increase of $f$ below $T = 32$ K. (b) The $^{75}$As NMR spectrum of the $Ca_{0.33}Na_{0.67}Fe_2As_2$ crystal at typical temperatures, under a 9.28 T magnetic field applied along the $c$-axis. The solid lines are gaussian fit to the spectrum.

Fig. 3: (color online) The temperature dependence of the $^{75}K$ (solid diamond) and the $^{23}K$ (solid triangle) of $Ca_{0.33}Na_{0.67}Fe_2As_2$ with field applied along the crystal $c$-axis. Hollow diamonds show the $^{75}K$ of $CaFe_2As_2$ (adapted from Ref. [19]) for comparison.

Fig. 4: (color online) Main panel: The spin-lattice relaxation rate $1/^{75}T_1$ at different temperatures, measured with field along the $c$-axis (circles) and along the $ab$-plane (diamonds). Inset: The log-log plot of $1/^{75}T_1$ vs. temperature with $H//c$. The solid line

is a guide to the power law $1/^{75}T_1 \sim T$.

Fig. 5: (color online) Main panel: The $^{23}$Na spectral linewidth. The solid line is a function fit to conventional *s*-wave superconductivity (see text). Inset: The $^{23}$Na NMR spectra at different temperatures. The solid lines represent Gaussian fits.

Fig. 6: (color online) Main panel: The $1/^{75}T_1T$ with two field orientations. The solid line is a fitting to $1/T_1T = A/(T+\Theta)$ (see text). The $1/^{75}T_1T$ of the parent compounds CaFe$_2$As$_2$ is also shown (diamonds) (adapted from Ref. [19]).

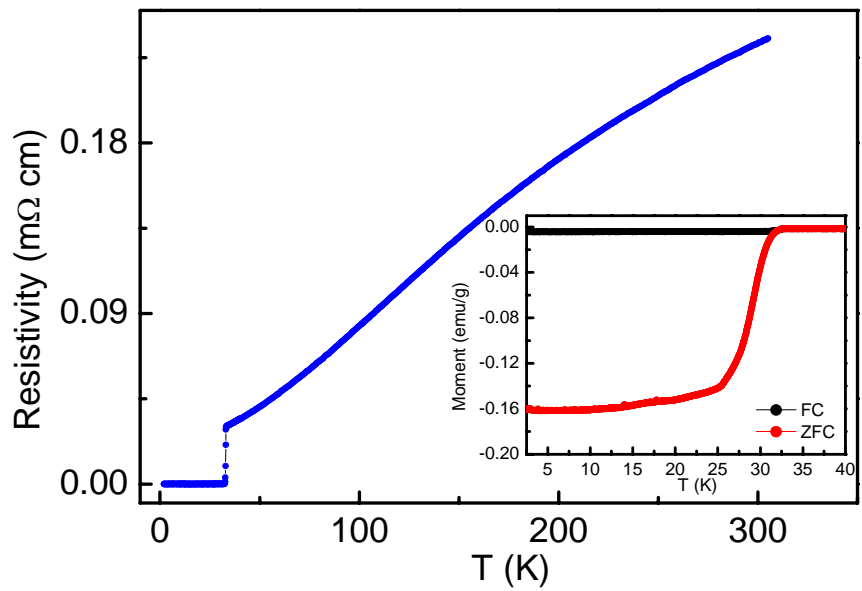

**Fig. 1**

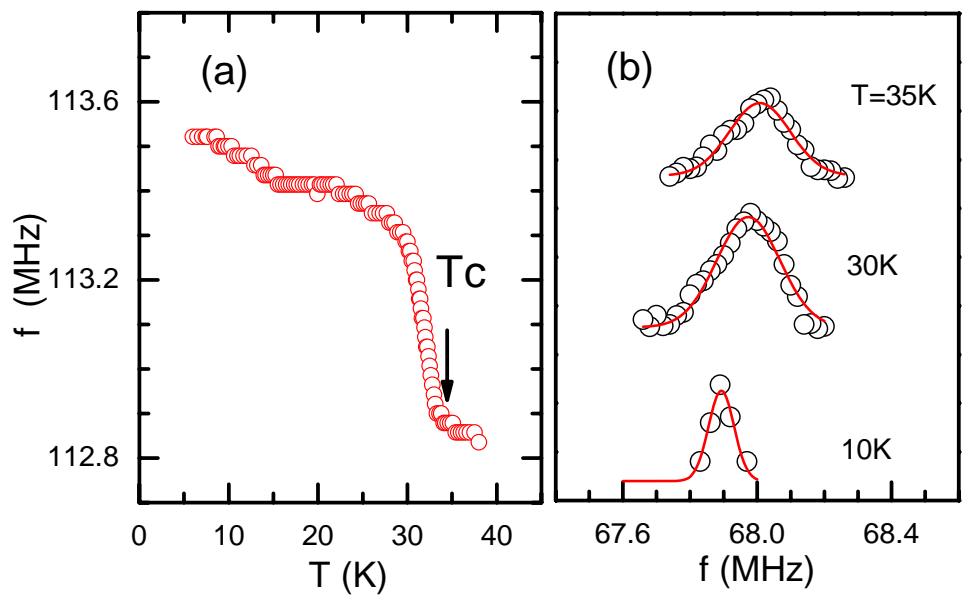

**Fig. 2**

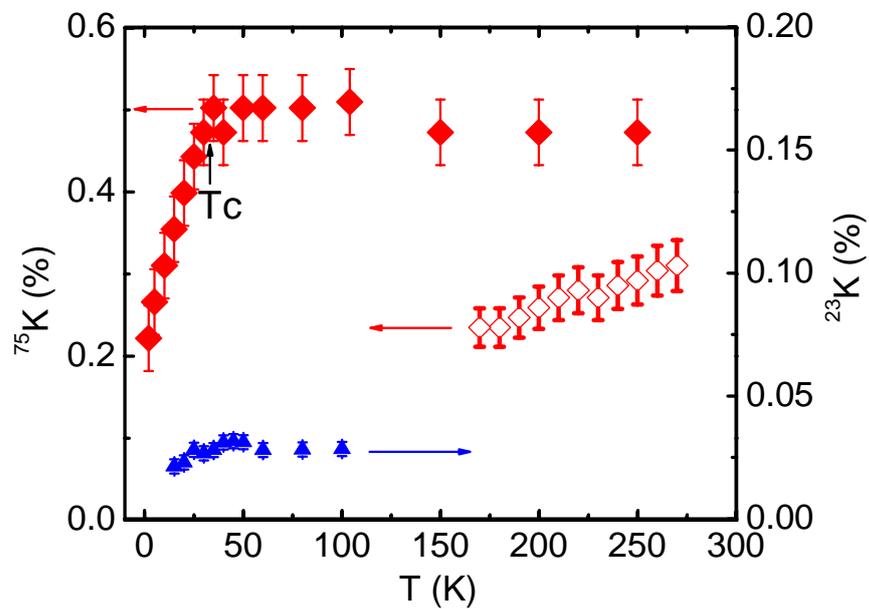

**Fig. 3**

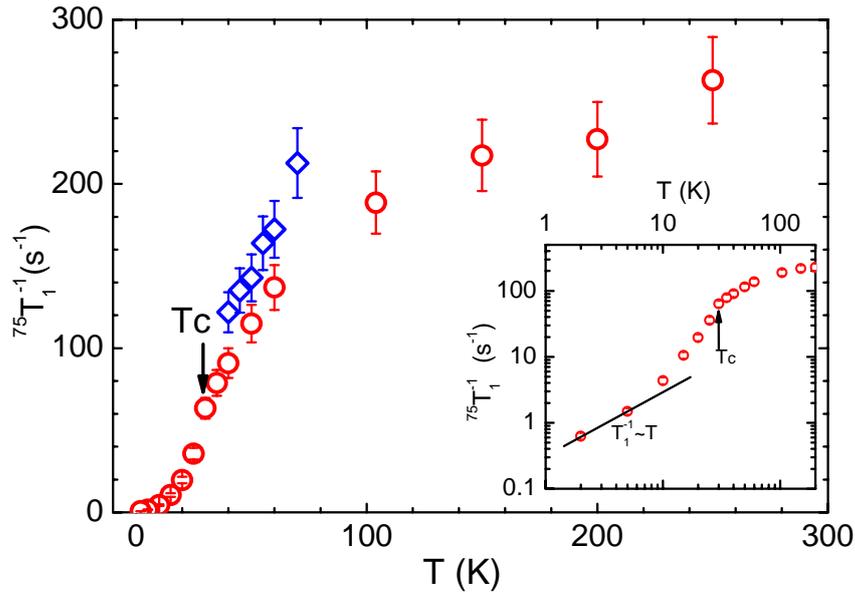

**Fig. 4**

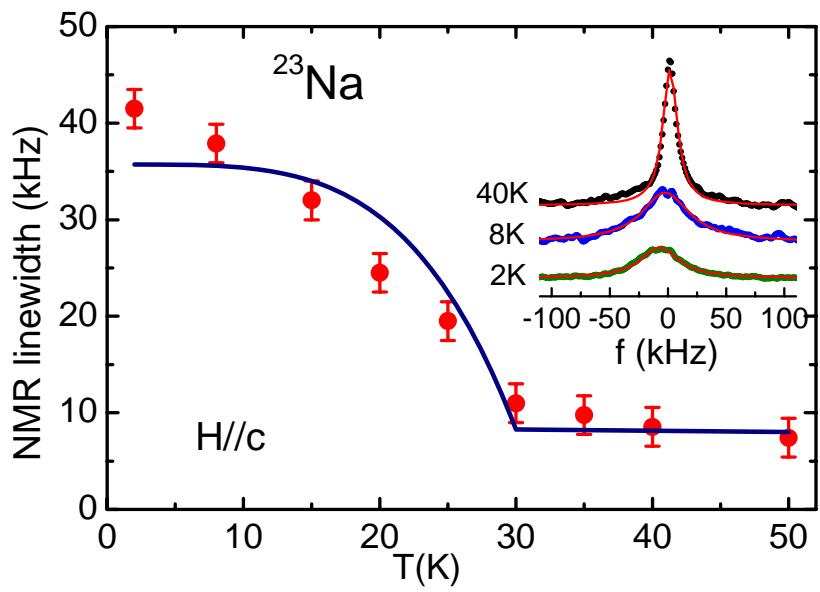

**Fig. 5**

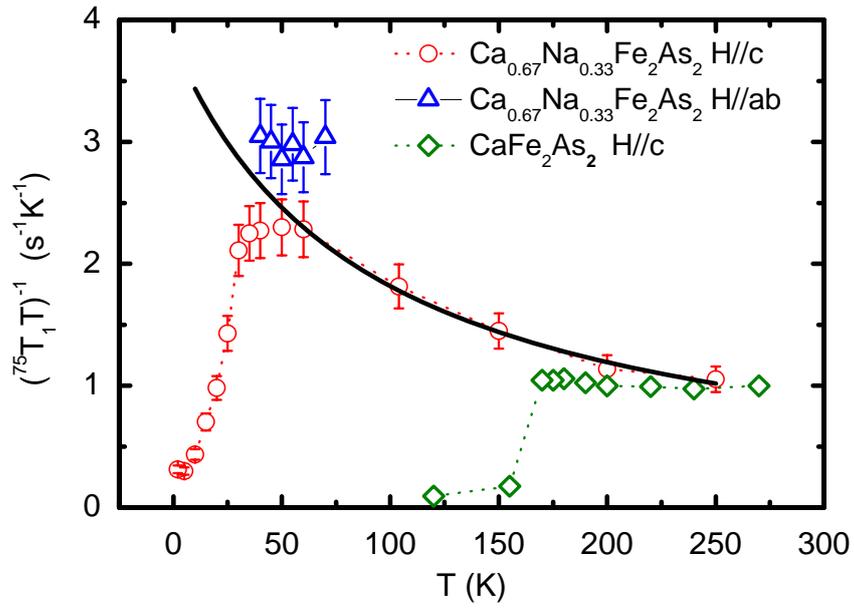

**Fig. 6**